\documentclass[12pt,epsf]{article}
\usepackage{amssymb,amsmath}
\usepackage{graphicx}

\newcommand{\be}{\begin{equation}}
\newcommand{\ee}{\end{equation}}
\newcommand{\bea}{\begin{eqnarray}}
\newcommand{\eea}{\end{eqnarray}}
\newcommand{\beas}{\begin{eqnarray*}}
\newcommand{\eeas}{\end{eqnarray*}}
\newcommand{\ba}{\begin{array}}
\newcommand{\ea}{\end{array}}

\newcommand{\tr}{\mathrm{Tr}}

\newcommand{\nbox}{{\,\lower0.9pt\vbox{\hrule \hbox{\vrule height 0.2 cm \hskip 0.19 cm \vrule height 0.2 cm}\hrule}\,}}

\def\href#1#2{#2}

\begin{document}
\begin{titlepage}
\hfill
\vbox{
    \halign{#\hfil         \cr
           } 
      }  
\vspace*{20mm}
\begin{center}
{\Large \bf Comments on the Bagger-Lambert theory  \\ and multiple M2-branes}

\vspace*{15mm}
\vspace*{1mm}
Mark Van Raamsdonk

\vspace*{1cm}

{Department of Physics and Astronomy,
University of British Columbia\\
6224 Agricultural Road,
Vancouver, B.C., V6T 1W9, Canada \\
e-mail: mav@phas.ubc.ca}

\vspace*{1cm}
\end{center}

\begin{abstract}
We study the $SO(8)$ superconformal theory proposed recently by Bagger and Lambert as a possible worldvolume theory for multiple M2-branes. For their explicit example with gauge group $SO(4)$, we rewrite the theory (originally formulated in terms of a three-algebra) as an ordinary $SU(2) \times SU(2)$ gauge theory with bifundamental matter. In this description, the parity invariance of the theory, required for a proper description of M2-branes, is clarified. We describe the subspace of scalar field configurations on which the potential vanishes, and note that this does not coincide with the moduli space for a stack of M2-branes. Finally, we point out a difficulty in constructing the required set of superconformal primary operators which should be present in the correct theory describing multiple M2-branes.

\end{abstract}

\end{titlepage}

\vskip 1cm

\section{Introduction}

In this note, we investigate a fascinating 2+1 dimensional field theory proposed recently by Bagger and Lambert \cite{bl1,bl2,bl3}  as a worldvolume description of multiple M2-branes in M-theory.\footnote{For a review of properties of M2-branes, see \cite{bermanreview}.} Like the ${\cal N}=4$ supersymmetric Yang-Mills theory in 3+1 dimensions, this theory has an explicit Lagrangian description, which the authors construct based on a new algebraic structure called a ``three-algebra,'' (an equivalent structure was proposed in \cite{gust}) reviewed in section 2 below. Bagger and Lambert have shown that one obtains an ${\cal N}=8$ supersymmetric theory with manifest $SO(8)$ R-symmetry given any such three-algebra, and argued that the theory must be superconformally invariant. So far, only a single example of such an algebra, given in \cite{bl2}, is known, and this example was proposed to describe the worldvolume theory of three M2-branes.

In this note, we explicitly rewrite the Bagger-Lambert theory in this explicit example as an ordinary gauge theory with gauge group $SU(2) \times SU(2)$ and matter in the bifundamental representation. This construction clarifies how the theory is able to maintain parity invariance - required if the theory is to describe M2-branes - despite the presence of a Chern-Simons term. Specifically, we find that twisted Chern-Simons term in the original formulation of the theory breaks up into separate Chern-Simons terms for the two $SU(2)$ gauge fields with opposite sign. While each of these is odd under parity, the combination is parity-invariant if we stipulate an exchange of the two gauge fields under parity.

We next revisit the moduli space for the theory, discussed previously in \cite{bl3}. For the explicit example, we find that the subspace of the vector space of scalar field configurations on which the potential vanishes is $(R^8 \times R^8)/O(2)$ where the $O(2)$ rotates the two $R^8$ factors into each other. On the other hand, the moduli space for $N$ M2-branes in uncompactified M-theory is $(R^8)^N/S_N$, which does not match our result for any value of $N$, even allowing for the addition of an extra decoupled center of mass factor to the Bagger-Lambert theory.\footnote{After the original preprint of this paper appeared, Jacques Distler pointed out \cite{distlerblog} that it is important to take into account the gauge field in determining the complete moduli space. See \cite{dmpv,lt} for a complete discussion.} We show that the $SU(2) \times SU(2)$ gauge symmetry is broken to $U(1)$ at a generic point on the moduli space, while there is a special vector subspace of the moduli space that preserves $SU(2)$.

Finally, we comment on gauge-invariant operators in the Bagger-Lambert theory.\footnote{This section of the paper arose from a discussion with Jaume Gomis, who suggested thinking about chiral operators in the Bagger-Lambert theory.} $AdS/CFT$ duality predicts that these theories should contain superconformal primary operators in traceless symmetric representations of $SO(8)$ with any number of indices (with the exception of the theory of two M2-branes, where due to the stringy exclusion principle, such representations with an odd number of indices are absent from the interacting part of the theory). On the other hand, we argue that the odd-index representations cannot be constructed from fields in the Bagger-Lambert theory, unless there is some additional algebraic structure (e.g. an ordinary product). Such structure is certainly not there in the $SO(4)$ example, so it is difficult to see how this case could provide a worldvolume theory for M2-branes, apart from possibly the $N=2$ case.

Despite the apparent conflicts between the Bagger-Lambert theory and our expectations for the M2-brane worldvolume theory, we feel that it is highly likely that the two are closely related. Perhaps the current example provided by Bagger and Lambert is analogous to ${\cal N}=4$ supersymmetric gauge theory for gauge group $G \ne U(N)$: this has all of the right symmetries but does not quite provide the right theory to describe stacks of D3-branes in noncompact type IIB string theory. Thus, the right M2-brane theory might arise from the Bagger-Lambert construction for a slightly more general algebraic structure. On the other hand it is possible that our analysis has simply been too naive, perhaps not correctly taking into account quantum effects or some nontrivial role for the gauge field.

Note: After this work was completed, the papers \cite{schwarz, mukhi, berman} appeared, which have some overlap with the present work.

\section{Review of the Bagger-Lambert construction}

To begin, we briefly recall the Bagger-Lambert construction of a class of $SO(8)$ superconformal theories. This starts by defining a three-algebra to be a vector space with positive definite inner product, together with a completely antisymmetric triple product, where the inner product and the triple product are defined by their action on a basis $T^a$ by
\[
\tr(T^a T^b) = h^{ab}
\]
and
\[
[T^a, T^b, T^c] = f^{abc} {}_d T^d \; .
\]
The triple product is required to satisfy
\[
[A,B,[C,D,E]] = [[A,B,C],D,E] + [C,[A,B,D],E] + [C,D,[A,B,E]] \; ,
\]
so that the operation $[A,B,*]$ behaves like a derivation when acting on a triple product of elements, and also
\[
\tr(A[B,C,D]) = - \tr([A,B,C],D)
\]
so that $f^{abcd} = h^{de} f^{abc} {}_e$ must be totally antisymmetric.

Given such an algebraic structure, one constructs a field theory starting with eight algebra-valued scalars $X^I_a$ transforming in the vector of $SO(8)$, eight algebra valued spinors transforming in the antichiral spinor representation of $SO(8)$, and a gauge field $A_{\mu ab}$ antisymmetric in the algebra indices. The spinors may be arranged into a single 32-component Weyl spinor $\Psi_a$, obeying
\[
\Gamma_{012} \Psi = - \Psi \; ,
\]
where we will use the notation $\Gamma^I$ to denote $32 \times 32$ Dirac matrices.
Defining
\[
\tilde{A}_\mu^c {}_d = f^{abc} {}_d A_{\mu ab}
\]
we can define covariant derivatives $(\tilde{D}_\mu X^I)_a$ and $(\tilde{D}_\mu \Psi)_a$ and a field strength $\tilde{F}_{\mu \nu}^a {}_b$ via the standard definitions, it may be checked that these transform covariantly  under a gauge-symmetry
\beas
\delta X^I_c &=& \tilde{\Lambda}^d {}_c X^I_d \cr
\delta \Psi_c &=& \tilde{\Lambda}^d {}_c \Psi_d \cr
\delta \tilde{A}_{\mu}^d {}_c &=& \tilde{D}_\mu \tilde{\Lambda}^d {}_c \cr
\tilde{\Lambda}^c {}_d &\equiv& f^{abc}{}_d \Lambda_{ab}\; .
\eeas
With these definitions, the Bagger-Lambert action is
\beas
{\cal L} &=& -{1 \over 2} D^\mu X^{Ia} D_\mu X^I_a + {i \over 2} \bar{\Psi}^a \Gamma^\mu D_\mu \Psi_a + {i \over 4} \bar{\Psi}_b \Gamma_{IJ} X^I_c X^J_d \Psi_a f^{abcd} \cr
&& - {1 \over 12} \tr([X^I,X^J,X^K][X^I,X^J,X^K]) \cr
&& + {1 \over 2} \epsilon^{\mu \nu \lambda} (f^{abcd} A_{\mu a b} \partial_\nu A_{\lambda c d}  + {2 \over 3} f^{cda} {}_g f^{efgb} A_{\mu ab} A_{\nu cd} A_{\lambda ef})
\eeas
In \cite{bl2}, this was shown to be invariant under gauge transformations and 16 supersymmetries:
\beas
\delta X^I_a &=& i \bar{\epsilon} \Gamma^I \Psi_a \cr
\delta \Psi_a &=& D_\mu X^I_a \Gamma^\mu \Gamma^I \epsilon - {1 \over 6} X^I_b X^J_c X^K_d f^{bcd} {}_a \Gamma^{IJK} \epsilon \cr
\delta \tilde{A}_\mu^b {}_a &=& i \bar{\epsilon} \Gamma_\mu \Gamma_I X^I_c \Psi_d f^{cdb} {}_a \; .
\eeas
where the spinor $\epsilon$ has the opposite chirality from $\Psi$,
\[
\Gamma_{012} \epsilon = \epsilon \; .
\]

\subsection{Example}

The only known example of this algebraic structure was given by Bagger and Lambert in \cite{bl2}. In this case, the vector space is $R^4$ and we can take
\beas
h^{ab} &=& \delta^{ab} \cr
f^{abcd} &=& f \epsilon^{abcd}
\eeas
for some constant $f$. In this case, the triple product is the natural generalization to four dimensions of the usual cross product: it gives a new vector perpendicular to the vectors in the product whose length is the signed volume of the parallelepiped spanned by the vectors.

\section{Description as a bifundamental gauge theory}

We will now see that for the known case just described, the Bagger-Lambert theory may be rewritten explicitly as an ordinary gauge theory with gauge group as $SU(2) \times SU(2)$, and matter in the bifundamental representation.

Under the  $SU(2) \times SU(2)$ decomposition, a real vector of $SO(4)$ becomes a bifundamental of $SU(2) \times SU(2)$, obeying the reality condition
\[
X_{\alpha \dot{\beta}} = \epsilon_{\alpha \beta} \epsilon_{\dot{\beta} \dot{\alpha}} (X^\dagger)^{\dot{\alpha} \beta}
\]
Explicitly, we can write
\[
X^{I} = {1 \over 2} \left( \ba{cc} x^I_4 + i x^I_3 & x^I_2 + i x^I_1 \cr -x^I_2 + i x^I_1 & x^I_4 - i x^I_3 \ea \right)
\]
with a similar expression for the spinor.

The gauge field $A_{\mu ab}$ may be decomposed into self-dual and anti-self-dual parts
\[
A_{\mu a b} = - {1 \over 2f} (A^+_{\mu a b} + A^-_{\mu a b}) \qquad \qquad A_{\mu a b}^\pm = \pm {1 \over 2} \epsilon_{abcd} A^\pm_{\mu cd}
\]
in terms of which we define
\[
A_\mu = A^+_{\mu 4 i} \sigma_i \qquad \qquad \hat{A}_\mu = A^-_{\mu 4 i} \sigma_i
\]
where the Pauli matrices $\sigma_i$ are normalized so that $\tr(\sigma_i \sigma_j) = 2 \delta_{ij}$. Making all the replacements, we find that the action becomes
\beas
{\cal L} &=&  \tr( -(D^\mu X^I)^\dagger D_\mu X^I + i \bar{\Psi}^\dagger \Gamma^{\mu} D_\mu \Psi )\cr
&& + \tr(-{2 \over 3} if \bar{\Psi}^\dagger \Gamma_{IJ} (X^I X^{J \dagger} \Psi + X^J \Psi^\dagger X^I + \Psi X^{I \dagger} X^J)  - {8 \over 3}f^2 X^{[I} X^{J \dagger} X^{K]}X^{K \dagger} X^J X^{I \dagger}) \cr
&& + {1 \over 2f} \epsilon^{\mu \nu \lambda} \tr(A_\mu \partial_\nu A_\lambda + {2 \over 3}i A_\mu A_\nu A_\lambda)  - {1 \over 2f} \epsilon^{\mu \nu \lambda} \tr(\hat{A}_\mu \partial_\nu \hat{A}_\lambda + {2 \over 3}i \hat{A}_\mu \hat{A}_\nu \hat{A}_\lambda) \cr
\eeas
where
\[
D_\mu X^I = \partial_\mu X^I + i A_\mu X^I -i X^I \hat{A}_\mu
\]
The supersymmetry transformation rules above become
\beas
\delta X^I &=& i \bar{\epsilon} \Gamma^I \Psi \cr
\delta \Psi &=& D_\mu X^I \Gamma^\mu \Gamma^I \epsilon +{2 \over 3} f X^I X^{J \dagger} X^K \Gamma^{IJK} \epsilon \cr
\delta A_\mu &=& f \bar{\epsilon} \Gamma_\mu \Gamma_I (X^I \Psi^\dagger - \Psi X^{I \dagger}) \cr
\delta \hat{A}_\mu &=& f \bar{\epsilon} \Gamma_\mu \Gamma_I (\Psi^\dagger X^I - X^{I \dagger} \Psi ) \; .
\eeas
Note that the twisted Chern-Simons term in the original formulation has decomposed into two separate ordinary Chern-Simons terms for $A$ and $\hat{A}$, albeit with opposite signs. The usual constraint that arises by demanding invariance under large gauge transformations then requires us to choose
\[
f = {2 \pi \over k}
\]
where the level $k$ is an integer.  It is particularly interesting to note that after a rescaling $A \to \sqrt{f}A$, all interaction terms in the theory are proportional to positive powers of $f$, so the theory becomes weakly coupled in the limit of large $k$. Thus, the theory can be solved exactly in the limit of large level and studied in perturbation theory for large level. The presence of the discrete parameter $k$ is a bit puzzling from the point of view of a possible M2-brane interpretation since it is not clear what this could correspond to, and in particular, we do not expect a weakly coupled limit of the theory. It seems important to understand the physical significance of this parameter.

\subsection{Parity invariance}

In the explicit expression for the action above it is straightforward to see how the theory manages to have a parity invariance symmetry despite the presence of Chern-Simons terms which are parity-odd.\footnote{Here, we take parity to be defined as a reflection in the $x_1$ direction.} Since the Chern-Simons terms for $A$ and $\hat{A}$ have opposite sign, the action of parity combined with the switch
\[
A_\mu \leftrightarrow \hat{A}_\mu
\]
leaves the gauge field part of the action invariant. In order to make the remainder of the bosonic action parity invariant, we must also demand a transformation
\[
X^I \leftrightarrow X^{I \dagger} \; .
\]
Invariance of the full action presumably now follows from supersymmetry, however we refer the reader to \cite{schwarz} for a more explicit discussion of parity for the fermionic terms in the action. Inspection of the kinetic term for the fermions shows that the correct transformation for these is
\[
\Psi \leftrightarrow \Gamma^1 \Psi^\dagger \; .
\]
where the $\Gamma^1$ factor comes from the standard parity transformation. All of these parity transformations may be seen to arise in the original language from a transformation that combines spacetime parity and a flip of the (234) directions in the internal space.

\section{Moduli Space}

In this section, we revisit the moduli space for the $SO(4)$ example of the Bagger-Lambert theory. We note that moduli space arising from the scalar fields does not quite match any of the known M2-brane moduli spaces.

To begin, we recall that in this case, the bosonic matter fields are 8 (distinguishable) vectors in an $R^4$ that is rotated by the gauge symmetry. The triple product gives a new vector perpendicular to the vectors in the product whose length is the signed volume of the parallelepiped spanned by the vectors. The bosonic potential is proportional the square of this volume, summed over each possible triple of vectors.

With this description, it is clear that the bosonic potential vanishes if and only if any three of the vectors lie in the same plane. This space is labeled by ordered sets of 8 vectors all of which lie in the same plane, with sets related by overall rotations in $R^4$ considered equivalent. Without loss of generality, we may assume that all vectors lie in the $x_3-x_4$ plane; the 8 $x_3$ coordinates and 8 $x_4$ coordinates form ordered octuplets which are rotated into each other by the residual $O(2)$ gauge symmetry. We conclude that the scalar moduli space for the $SO(4)$ Bagger-Lambert theory is $(R^8 \times R^8) / O(2)$.\footnote{Our result differs from the one in \cite{bl3} by the presence of the $O(2)$ factor. }

For $N$ M2-branes in uncompacified M-theory, the moduli space should be simply $(R^8)^N/S_N$. This does not agree with the space of scalar fields with vanishing potential for any value of $N$, even allowing for the addition of a decoupled center-of-mass sector of the theory as was done in \cite{bl3}.\footnote{As pointed out by Jacques Distler \cite{distlerblog} after the original preprint of this paper appeared, the space of scalar fields with vanishing potential is not quite the full moduli space of the theory, since it is important to take into account the gauge field. Nevertheless, the full answer \cite{dmpv,lt} still does not agree with what is expected for M2-branes in flat space.}

We briefly comment on the symmetry-breaking structure of the moduli space. In our bifundamental notation, the moduli space turns out to be exactly the set of matrices $X^I$ that are diagonal up to gauge transformations. A generic point on the moduli space may be described by a matrix
\[
X^{I} =  \left( \ba{cc} z^I & 0 \cr 0 & \bar{z}^I \ea \right)
\]
where $z^I$ are complex. This preserves residual $U(1)$ gauge symmetry, generated by $A_\mu = \hat{A}_\mu \propto \sigma_3$. On the vector subspace of the moduli space where $z^I$ is real, a full $SU(2)$ is preserved, generated by $A_\mu = \hat{A}_\mu$.

It should be interesting to expand the action explicitly about generic and non-generic points on the moduli space.

\section{Superconformal Operators}

The correct superconformal theory describing multiple M2-branes is believed to be dual to M-theory on $AdS^4 \times S^7$, with the curvature of the spacetime in Planck units determined by the number $N$ of M2-branes \cite{maldacena}. For large $N$, the curvature is small, and supergravity should provide a good description of the low energy physics. Thus, low-dimension operators in the superconformal field theory should be in one-to-one correspondence with the spectrum of supergravity fluctuations around the $AdS^4 \times S^7$ background. The single-particle states were determined in \cite{cenm} and shown in \cite{gw,shiraz,ofer} to correspond to a single series of irreducible representations of the superconformal algebra, labeled by an integers $k \ge 1$. The operators of lowest dimension in each of these representations, are superconformal primary operators of dimension $k/2$ transforming in the symmetric traceless k-index representation of the $SO(8)$ R-symmetry group. Thus, such operators should be present in the conformal field theory that describes the decoupled physics of a large number of M2-branes.

In the Bagger-Lambert theory for a general 3-algebra, the matter fields $X^I_a$ and $(\Psi_\alpha)_a$ transform in the $8_v$ and $8_c$ representations of $SO(8)$ and carry a single algebra index (they are elements of the algebra itself). Meanwhile, the gauge fields are $SO(8)$ invariant and carry two algebra indices. To form gauge invariant operators, all algebra indices must be contracted. In the absence of any additional algebraic structure, the only invariant tensors that we have to work with are $h^{ab}$ and $f^{abcd}$ (this is certainly true in the $SO(4)$ example). As a result, all gauge-invariant operators must have an even number of matter fields. The tensor product of two $8_c$ representations gives representations appearing in the tensor product of even numbers of $8_v$ representations. Thus, the $SO(8)$ representation of any bosonic gauge-invariant operator must be an ordinary tensor representation of $SO(8)$ with an even number of indices. In particular, it seems impossible to construct the expected operators in symmetric, traceless representations of $SO(8)$ with an odd number of indices.

In light of the observations in the previous paragraph, it is puzzling how the $SO(4)$ theory we have focused on this paper could provide the worldvolume theory for some number of M2-branes.\footnote{Shiraz Minwalla has emphasized that the moduli space and the operator spectrum are intimately linked \cite{shiraz2}, so this may be related to our observations in section 4.} For the special case of two M2-branes, the situation is slightly better, due to the stringy exclusion principle \cite{mst,ms,hrt,jr}, which reduces the expected spectrum of operators from the full supergravity result. In this case, the expected result for the operator spectrum (conjectured in (\cite{shiraz2}) for the interacting part of the M2-brane theory does not contain symmetric traceless representation with an odd number of $SO(8)$ indices.\footnote{This may be seen by dividing $Z_2^2$ by $Z_2^1$ in equation (3.25) \cite{shiraz2} and noting the absence of terms $x_k^{2n+1}$.} However, the moduli space issue is still problematic for this case.

For other possible examples of three-algebras, our discussion above shows that if they are to describe worldvolume theories of M2-branes, there must be some additional algebraic structure that allows us to form gauge-invariant operators more general than those constructed from the invariants $h^{ab}$ and $f^{abcd}$ alone.

\section*{Acknowledgements}

I am grateful to Jim Bryan, Jaume Gomis and Shiraz Minwalla for helpful discussions, and to Jacques Distler for comments on the original version of this manuscript. This work has been supported in part by the Natural Sciences and Engineering Research Council of Canada, the Alfred P. Sloan Foundation, and the Canada Research Chairs programme.

\end{document}